\documentclass[sigconf,nonacm]{acmart}

\makeatletter
\renewcommand{\@ACM@checkaffil}{}
\makeatother

\AtBeginDocument{%
  }

\usepackage{tikz}
\usepackage{todonotes}
\usepackage{listings}
\usepackage{booktabs}
\usepackage{tabularx}
\usepackage{array}
\usepackage{float}
\newfloat{listing}{tb}{lol}
\floatname{listing}{Listing}
\usepackage{fvextra}
\input{sections/pygstyle}
\fvset{fontsize=\small}

\renewcommand\footnotetextcopyrightpermission[1]{}

\begin{document}

\title{No, Your Tools Are Not Enough: It's Time to Bring Agents to the Hunt}

\author{Ati Priya Bajaj}
\authornote{Equal contribution}
\affiliation{%
  \institution{Arizona State University}
}
\email{atipriya@asu.edu}

\author{Xiang Mei}
\authornotemark[1]
\affiliation{%
  \institution{Arizona State University}
}
\email{n132@asu.edu}

\author{Tiffany Bao}
\affiliation{%
  \institution{Arizona State University}
}
\email{tbao@asu.edu}

\author{Yan Shoshitaishvili}
\affiliation{%
  \institution{Arizona State University}
}
\email{yans@asu.edu}

\author{Ruoyu Wang}
\affiliation{%
  \institution{Arizona State University}
}
\email{fishw@asu.edu}

\author{Adam Doupé}
\affiliation{%
  \institution{Arizona State University}
}
\email{doupe@asu.edu}

\begin{abstract}
  Modern software systems are large and complex, demanding scalable and autonomous techniques for vulnerability discovery.
Existing approaches still depend on human expertise for domain knowledge, and on human intervention throughout the process of vulnerability discovery.
AI agents offer us an opportunity to make vulnerability discovery more adaptive, scalable, and autonomous.
Agents bring the ability to make continuous decisions and adapt, and general knowledge that can be augmented.

In this paper, we present our experience using AI agents to find vulnerabilities across diverse programs such as Windows drivers, the Linux kernel, XNU, firmware, and web applications.
We use an iterative agentic workflow and find that it generalizes across these targets.
We discuss tasks that agents help with, which were previously challenging, and some future research directions we consider important for advancing autonomous vulnerability discovery.
\end{abstract}


\maketitle

\makeatletter
\fancyhead[RE]{}
\fancyhead[LO]{\ACM@linecountL}
\makeatother

\section{Introduction}\label{sec:intro}

Programs underpin our modern world.
Our lives are so intertwined with programmed devices around us, such as cars, smartwatches, and wearable medical devices, that any critical vulnerability in them can interfere with or even interrupt everyday operations, expose sensitive information, cause financial losses, or even harm human lives.

In 2025, 48,244 CVEs were reported ~\cite{cveMetrics2025}.
On average, a security breach costs 4.44 million dollars ~\cite{ibm2025dataBreachReport} and takes approximately 241 days to identify and contain.
As of 2026, approximately 31\% of breaches start with software vulnerability ~\cite{verizon2026dbirPress}.
For example, the ransomware attack on the MOVEit Transfer web application in 2023 exploited a zero-day SQL injection vulnerability~\cite{cisa-moveit, mandiant-moveit} and gained unauthorized access to the database.
Because this app is used to transfer sensitive data, the breach exposed personally identifiable information (PII) and protected health information (PHI) of millions of individuals~\cite{emsisoft-moveit, sternsecurity-moveit} across 2,000 organizations in the healthcare, education, finance, and government sectors, with an estimated loss of 400 million dollars in the health sector alone.

A vulnerability in a widely used library can have a significant impact when it is used as a direct or transitive dependency in other software.
For example, Apache Log4j, a Java logging library, dynamically resolves string placeholders in log messages via lookups.
In 2021 Log4Shell~\cite{log4shell} exploited this feature, and an unauthenticated user could perform remote code execution (RCE) on a production server.
This impacted approximately 17k Java packages~\cite{google-log4j}.

Similarly, OpenSynergy's BlueSDK, a Bluetooth SDK used in more than one billion IoT devices and 350 million cars, was found to be vulnerable to memory corruption and logic bugs leading to a 1-click RCE ~\cite{perfektblue}.
An attacker could track GPS, record audio, or move laterally into the system after a single user action that enabled the attack.

With so many software systems out there, manually finding these vulnerabilities is intractable.
Therefore, security researchers, practitioners, and government have spent decades pushing toward more scalable and autonomous vulnerability discovery.
In 2016, DARPA launched the Cyber Grand Challenge to automatically find vulnerabilities in software programs, generate proofs of concept (PoCs), and patch them~\cite{cgc2016}.
Every technique has its own limitations, so to counter that, teams combined techniques to push the technical and scientific boundary of vulnerability research~\cite{driller2016}.

Eight years later, DARPA began exploring the use of large language models (LLMs)~\cite{gpt2} for securing software and launched AIxCC, moving the field another step closer to autonomous vulnerability discovery and mitigation.
At the beginning, the models were not nearly as capable as they are today, so participants relied more on traditional techniques than on AI.
By the finals in August 2025, the models improved, and the finalist systems incorporated much more AI and found 18 zero-day vulnerabilities in production open-source software~\cite{aixcc2025}.

Although the security community is increasingly recognizing the potential of AI, especially LLMs, for vulnerability discovery, few have publicly presented an end-to-end workflow that demonstrates how agentic approaches can effectively apply to vulnerability discovery across diverse domains.
In this paper, we aim to bridge this gap by sharing our workflows and practical experience using AI agents for real-world vulnerability discovery in a diverse set of targets including operating system kernels, firmware, Windows drivers, and web applications (\S\ref{sec:domains}).
Our experience demonstrates that currently available general-purpose LLMs, despite lacking the security specialization of models such as Mythos and Fable 5, are already capable of enabling effective vulnerability discovery.
More importantly, we show \emph{how} vulnerability researchers can systematically use these readily available models in practice.

We find that agents are most effective in a multi-step, structured workflow rather than being applied one-shot to find vulnerabilities in a program.
Well-scoped vulnerability analysis along with dynamic verification can significantly reduce false positive reports generated by the agents.
We also observe that the quality of reverse-engineering tools and program representations can enhance the performance of agents.

In Section \ref{sec:whatnow}, we discuss the challenges and limitations that we encountered when deploying agentic workflows for vulnerability research and highlight several promising directions for future research.
Note that our work only focuses on vulnerability discovery and program analysis.
We neither investigate nor automate exploit generation, and our workflows are \emph{not} designed for weaponizing any vulnerabilities.

\section{Background}\label{sec:background}
We present a brief overview of the techniques used for vulnerability discovery.

\subsection{Static Vulnerability Discovery}\label{sec:static}

Static analysis is reasoning about code without executing it.
It is done over the abstract representation of the program such as control flow graphs (CFGs), data flow graphs (DFGs) or abstract syntax trees (ASTs).
By modeling program behavior in such a form, it can track data flow, state transitions or relationships between different components.
Static analysis is an over-approximation of program behavior.

There are different techniques for doing static analysis.
For example, taint analysis~\cite{saxena2008efficient} tracks how attacker-controlled data propagates through a program.
Some tools combine multiple program representations into a single representation or express them as a relational database that can be queried for analysis ~\cite{yamaguchi2014cpg, codeql}.
Some tools rely on pattern matching to find known vulnerable patterns in code ~\cite{semgrep}.
Static analysis is used to find vulnerabilities in different programs such as web applications, browsers, firmware, and operating systems~\cite{livshits2005java, jovanovic2006pixy, arzt2014flowdroid, brown2020sys, angrsok2016}.

\subsection{Dynamic Vulnerability Discovery}\label{sec:dynamic}

Dynamic analysis is observing the behavior of a program by executing it.
Some programs can run directly on the host machine, while others require a specific environment or hardware, so they are emulated for analysis.
Fuzzers ~\cite{miller1990, afl2014, fioraldi2020aflpp, libfuzzer, windowsfuzzer} run a program with randomly generated inputs (and mutate them) to find crashes or unexpected behavior.
Dynamic symbolic execution (concolic execution) starts with concrete input and uses constraint solvers to generate more inputs that explore different program paths.
Fuzzing is scalable but may not explore all program paths, while symbolic execution is precise but does not scale.
Dynamic analysis can be combined with static analysis to create hybrid approaches~\cite{driller2016, qsym2018} to balance the trade-offs between scalability and precision.

Another form of dynamic analysis is to emulate a program and observe its behavior.
This is especially useful for embedded firmware and closed-source programs that cannot be easily fuzzed.
Commonly used emulators are QEMU~\cite{qemu} and Unicorn~\cite{unicorn}.

\subsection{Machine-Learning-Based Vulnerability Discovery}\label{sec:ai-assisted}

Machine learning has been applied to vulnerability discovery for nearly a decade.
Early work explored different deep-learning architectures for vulnerability detection, such as recurrent models (RNNs), bidirectional Long Short-Term Memory (LSTMs), and graph neural networks (GNNs)~\cite{li2018vuldeepecker, li2022sysevr, zhou2019devign, li2021vuldeelocator}.
The models were trained on different program representations such as ASTs, CFGs, DFGs or even code fragments to capture program behavior and predict whether code is vulnerable, identify vulnerability types, or rank potentially vulnerable functions.

With the introduction of LLMs, vulnerability detection research shifted to using big pretrained language models.
These models are trained on large corpora of natural language and source code to capture knowledge that is transferable to various downstream tasks ~\cite{hanif2022vulberta, fu2022linevul}.
They can capture contextual relationships in code more effectively than many earlier deep-learning models.

Earlier approaches treated vulnerability detection as a prediction task, such as predicting the probability of each line being vulnerable or classifying Common Weakness Enumeration (CWE) types.
LLMs such as GPT-3~\cite{brown2020gpt3,jiang2023mistral,yang2025qwen3} are generative models that produce richer outputs such as vulnerable code snippets or vulnerability descriptions in natural language.
They have been used independently or alongside static and dynamic analysis tools, for tasks such as filtering out false positives or generating seeds ~\cite{li2025iris, llift2024, xia2024fuzz4all, chatafl2024}.

\subsection{Agentic Vulnerability Discovery}\label{sec:agentic}

An AI agent is a system that can autonomously reason, plan, perform actions and adapt based on feedback.
Modern AI agents use LLMs as their reasoning engine, have tools to perform actions, and memory to save interactions.
Agents can read code, form hypotheses, invoke tools, access memory and refine their strategy based on feedback.
This adaptive and iterative approach differentiates agentic vulnerability discovery from earlier machine-learning-based approaches.
Agents can have different objectives, such as finding vulnerability, generating proof of concept (PoC), or validating vulnerability.
Agents have already been used to find zero-day vulnerabilities and reproduce known CVEs in real-world systems~\cite{anthropic2026mythos,xbow,exploitbench, exploitgym}.

\section{Workflow for Agentic Vulnerability Discovery}\label{sec:workflow}

\begin{figure*}[t]
    \centering
    \includegraphics[width=\textwidth]{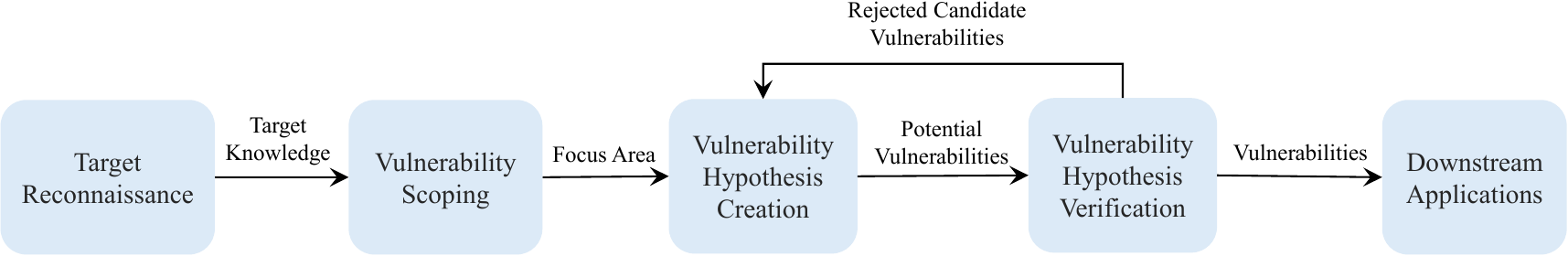}
    \caption{A high-level agentic vulnerability discovery workflow.}
    \label{fig:workflow}
\end{figure*}

Workflows for agentic vulnerability discovery follow a similar high-level structure, even though the software targets and downstream applications may differ.
This workflow can be divided into five phases, as shown in Figure~\ref{fig:workflow}.

\begin{enumerate}
\item \textbf{Reconnaissance.} The first step is to understand the target.
This may include identifying the software's entry points, the different components of the target software and how they interact, and identifying other open-source software or libraries used in the target.
The reconnaissance can be specific or broad depending on the human analyst's use case.

\item \textbf{Vulnerability Scoping.}
After building an understanding of the target, the next step is to determine which vulnerability classes are most relevant and which target components should be prioritized for further analysis.
For instance, if the target accepts and processes user-controlled input, the analysis may focus on the code that parses that input and on how the resulting data is used throughout the system.
Similarly, if the target reuses existing open-source libraries, the analysis may examine the versions of those libraries and whether any existing vulnerabilities also affect the target.

\item \textbf{Vulnerability Hypothesis Creation.}
After the scope of the analysis has been narrowed, the next step is to hypothesize potential vulnerabilities based on the understanding developed in the earlier steps.
These candidate vulnerabilities can be generated solely by the agent or combined with the output of existing tools such as fuzzers or taint analysis engines.
The agent may formulate concrete inputs or generate scripts that reach the vulnerable code.

\item \textbf{Vulnerability Hypothesis Verification.}
The potential vulnerabilities identified in the previous phase are validated.
They can be verified by running the generated inputs or scripts against the target or by using another agent to validate the findings.
If the candidate vulnerability is a false positive and fails the verification, it can be discarded or refined by gathering more evidence and then revalidating it.

\item \textbf{Downstream Applications.}
Successfully verified vulnerabilities can then be reported, analyzed to determine their root cause, patched, or combined with other vulnerabilities in the software to amplify their impact.
\end{enumerate}

\section{A Tour of the Evidence}\label{sec:domains}
In this section, we present our findings and experiences with using agentic workflows for vulnerability discovery across a diverse set of program targets.
\subsection{Linux Kernel}\label{sec:domain-kernel}


The Linux kernel is one of the largest and most thoroughly tested open-source projects.
It is widely deployed across web servers, mobile devices, and cloud infrastructure.
The codebase exceeds 30 million lines of C and is maintained by thousands of developers under a rigorous review process: patches go through subsystem maintainers, mailing-list review, and integration trees before reaching a release.
It has been continuously fuzzed by Google's syzkaller since 2016, stress-tested by sanitizers (KASAN, KMSAN, KCSAN), and analyzed by both commercial and academic static analysis tools.
Using LLM-assisted methods, we found around 200 unique crashes in the Linux kernel; more than 20 bugs have been assigned CVE numbers, and we are still disclosing the rest.
Below we walk through one of them, CVE-2026-31419, which we found, exploited, and patched, and which won a \$10{,}000 bounty on Google's kernelCTF~\cite{kernelctf}.

We found the double free in the bonding subsystem's broadcast transmit path, \texttt{bond\_xmit\_\allowbreak{}broadcast()}, using LLM-assisted fuzzing. The race is summarized in Figure~\ref{fig:bonding-race}.
The root cause is a race condition: the function iterates over a pre-snapshot (RCU) slave array but calls \texttt{bond\_is\_last\_slave()} to decide whether to reuse the original \texttt{skb} (zero-copy) or clone it.
However, \texttt{bond\_is\_last\_slave()} reads the tail of the linked list, and a slave enslave or release can modify that linked list concurrently.
Consider a bond with slaves \texttt{[A, B]}: the TX thread iterates A (not last, clones) and then B, which is the tail, so \texttt{bond\_is\_last\_slave()} returns true and the original \texttt{skb} is consumed (freed). Before the loop ends, a second thread enslaves C, shifting the tail to C. The TX thread continues to C, which is now also the tail, so \texttt{bond\_is\_last\_slave()} returns true again and the same freed \texttt{skb} is reused, producing a use-after-free.

\begin{figure}[h]
\centering
\small
\begin{tabular}{@{}r|l|l@{}}
\multicolumn{1}{c}{} & \multicolumn{1}{c}{\textbf{Thread 1 (TX)}} & \multicolumn{1}{c}{\textbf{Thread 2 (Config)}} \\ \hline
T1 & iter A: {\ttfamily is\_last}=false, clone & \\
T2 & iter B: {\ttfamily is\_last}=true, skb freed & \\
T3 & & enslave C; tail: B$\to$C \\
T4 & iter C: {\ttfamily is\_last}=true, \textcolor{red}{\textbf{UAF}} & \\
\end{tabular}
\caption{CVE-2026-31419 Race Condition. Thread~1 frees the skb at B (the tail). Thread~2 enslaves C, moving the tail. Thread~1 continues to C, which also passes \texttt{bond\_is\_last\_slave()}, reusing the freed skb.}
\label{fig:bonding-race}
\end{figure}

The fix~\cite{mei2026bonding} is a one-line change (Listing~\ref{lst:bonding-patch}): replace the racy predicate with an index comparison against a snapshot of the slave count taken before the loop.

\begin{listing}[h]
\centering
\resizebox{\columnwidth}{!}{%
\begin{minipage}{\columnwidth}
\ttfamily\small
\textcolor{gray}{// drivers/net/bonding/bond\_main.c} \\
\textcolor{red}{- if (bond\_is\_last\_slave(bond, slave)) \{} \\
\textcolor{green!50!black}{+ if (i + 1 == slaves\_count) \{} \\
\mbox{}\quad\quad skb2 = skb; \\
\mbox{}\quad\quad skb\_used = true; \\
\mbox{}\quad\} else \{
\end{minipage}%
}
\caption{CVE-2026-31419 Patch: replace the racy \texttt{bond\_is\_last\_slave()} predicate with a stable index comparison.}
\label{lst:bonding-patch}
\end{listing}

This bug survived for over four years (January 2022 to March 2026) despite continuous fuzzing and multiple rounds of code review.
Coverage-guided fuzzers do not spend resources on code paths that stop producing new coverage, and triggering this bug requires two threads reaching very different code paths in the same subsystem (a data path and a configuration path) to interleave in a tight window, which produces no positive coverage signal until the race actually fires.
Manual audit is similarly constrained: the kernel has over 30 million lines of code, and the number of security experts who review concurrency issues is small relative to the codebase.
An AI agent can link relevant functions automatically, identifying a cluster of functions across the data and configuration paths, and directing the fuzzer to execute and mutate the related corpus more often (using syzkaller's focus-area feature) to narrow the search space and eventually trigger the vulnerability.

Even after the fuzzer produced a crash, triaging a race condition bug remains difficult: the crash report shows a use-after-free in \texttt{skb\_clone} without a stable reproducer, and the root cause is not obvious from the stack trace alone.
The agent parsed the crash report, hypothesized the race condition as the root cause, and wrote a proof-of-concept to verify it. It failed many times, but completed the triage in several hours without manual effort.

For exploitation, we combined private and public exploitation techniques as a knowledge base for the LLM and actively guided it through the harder steps.
The double-free immediately triggers the kernel's hardening check, killing the process before the freed object can be reclaimed.
The agent spent two days attempting to work around this and kept failing.
A human expert analyzed the code and connected a \texttt{netem} scheduler (a completely unrelated networking submodule) to delay the second free, creating a window to reclaim the freed object and convert the double-free into a usable use-after-free primitive.
From that point, the agent completed the remaining exploitation steps: using the UAF primitive to achieve arbitrary code execution on a modern Linux distribution with KASLR, SMEP/SMAP, and SLAB hardening enabled.

The agent tries relevant public techniques but cannot tell whether a primitive fits until it encounters an error, while a trained expert can rule out mismatches upfront.
For example, to reclaim the freed skb and avoid an early crash, it attempted \texttt{msg\_msg} (now allocated from its own dedicated bucket) and \texttt{pipe\_\allowbreak{}buffer} (allocated with \texttt{GFP\_\allowbreak{}KERNEL\_\allowbreak{}ACCOUNT}, a different bucket from the target \texttt{skb}).
After we clarified the allocator state and asked it to use \texttt{sk\_buff} allocation to reclaim the freed \texttt{skb}, it started doing the right thing. The human expert helped the agent select the appropriate primitives and method for the final exploit to win the kernelCTF.
A recent report showed an LLM autonomously exploiting FreeBSD~\cite{carlini2026freebsd}, but FreeBSD ships without stack canaries and without KASLR by default, so exploitation there requires little beyond a working memory corruption primitive. On a hardened Linux kernel, as our experience shows, human expertise remains necessary for complex vulnerabilities at critical steps.

\begin{table}[H]
\centering
\small
\setlength{\tabcolsep}{4pt}
\renewcommand{\arraystretch}{1.15}
\caption{Approaches and models used across different software targets.}
\label{tab:domain-techniques-models}
\begin{tabularx}{\columnwidth}{@{}>{\raggedright\arraybackslash}p{0.28\columnwidth}>{\raggedright\arraybackslash}X>{\raggedright\arraybackslash}X@{}}
\toprule
\textbf{Target} & \textbf{Techniques} & \textbf{Models} \\
\midrule
Agentic Reimplementation & - & Various GPT models (April 2026)  \\
Firmware & Decompilation & GPT-5.4, Claude Opus 4.6 \\
Linux Kernel & Fuzzing, KASAN &  Claude Opus 4.6  \\
Web Applications & - & GPT-5.4, Claude Opus 4.6 \\
Windows Drivers & Decompilation, DSE & Claude Opus 4.6--4.8 \\
XNU & Decompilation, KASAN &  Claude Opus 4.6 \\
\bottomrule
\end{tabularx}
\end{table}

\subsection{XNU}\label{sec:domain-xnu}

XNU (``X is Not Unix'') is the operating system that underpins almost all products developed by Apple Inc., including iOS, iPadOS, WatchOS, visionOS, and tvOS.
Thus, XNU forms the core of billions of devices, and the security of XNU is critical.
In fact, Apple will pay up to \$2M USD for zero-click remote exploitation (to say nothing of the value of these on the grey or black markets).

What many people might not know is that XNU is open source~\cite{xnu-source} (the core of the OS is open source, but drivers and other modules are closed source).
Therefore, we used our agentic LLM-based workflow to find vulnerabilities in the open source XNU.
In this case, we prompted with a cross-product of targets (modules to focus on) and bug class (buffer OOB, use-after-free, etc.).
Then, our validation component used an XNU KASAN kernel to validate the potential discovered vulnerabilities. 

This resulted in several bugs, some of which were already fixed (the open-source XNU release lags behind the deployed version) and some which were 0-days and reported (and we discuss these where we can).

\subsubsection{Heap Overflow In \texttt{recvmsg\_x()} UIO Reuse}

\begin{listing}[tb]
  \input{sections/recvmsg_vuln.pygtex}
  \caption{A vulnerable code snippet of \texttt{recvmsg\_x} in XNU.}
  \label{lst:recvmsg}
\end{listing}

The LLM discovered a bug in the \texttt{recvmsg\_x} syscall.
This syscall is a batched version of \texttt{recvmsg}, to receive multiple datagrams in a single syscall.
While it is a private syscall intended for Apple frameworks, it is still accessible by unprivileged user code.

The vulnerability, shown in Listing~\ref{lst:recvmsg}, is a trivial inverted check in sending data back from the socket to the user.
This code uses \texttt{uio} structs to copy data back to the user, and the comment before the check specifically says to reuse the uio vector if it is large enough.
However, the check is checking if there \texttt{uio} has less than or equal to space, then it's reused.
This vulnerability causes a heap overwrite, and can be triggered by an unprivileged process.

This vulnerability was first introduced in XNU 10002.1.13 which shipped on MacOS as 14.0 (Sonoma) on September 26th, 2023, and was fixed in macOS 26.4 (Tahoe) on March 24th, 2026.
Interestingly, this vulnerability is trivial to identify when looking at the vulnerable code (and knowing what to look for), however it would be much more difficult to detect with fuzzing, as it would require a harness that calls \texttt{recvmsg\_x} with multiple sized buffers while also allowing the fuzzer to control the size and amount of the datagrams received on the network.

\subsubsection{Under Submission Bugs}

A number of vulnerabilities are currently under submission, and here we discuss the high-level details to highlight interesting findings.

One vulnerability that the system found was in a closed-source driver.
Here, we gave the LLM the ability to access IDA decompilation via MCP, and tasked it to again find different vulnerability classes in the driver.
One of these resulted in a reproducible use-after-free in a closed source driver.
This shows that the systems are able to find real vulnerabilities even in closed-source systems.

In fact, in an accidental experiment, the prompt asked the agentic AI to use the MCP server to analyze the driver but IDA was not running.
The model ended up using objdump to disassemble the driver (and still found issues).

Another interesting under submission vulnerability is a remotely triggerable use-after-free in a complex protocol (CVE-2026-28982).
In order to prove that this was a real vulnerability, the agentic LLM had to create a POC, and this involved creating one side of the network protocol (with XNU as the other side).
This was required to drive the protocol into the state to trigger the UAF, and this shows why fuzzing was unlikely to find the vulnerability.
Off-the-shelf open source implementations or libraries of protocols are difficult to use in this manner, because they are not meant for the flexibility of triggering vulnerabilities, so the agentic AI wrote its own implementation.
This bug was present for over a decade. 

\subsection{Windows Drivers}\label{sec:domain-windows}

We identified multiple vulnerabilities in Windows drivers for Windows 10 and 11, both in Microsoft's own drivers and in some third-party drivers, using mostly agentic vulnerability discovery with the help of static analysis (i.e., without fuzzing).
Finding vulnerabilities in Windows drivers presents a unique challenge:
Source code is not available for most Windows drivers, and debug symbols are only available if the analysis target is a first-party Microsoft Windows driver.
Unsurprisingly, we empirically found that LLMs are not effective at finding vulnerabilities in binary code without first converting it into a more human-readable format, e.g., C pseudo-code generated by a decompiler.
We do not disclose the details of these vulnerabilities in this manuscript due to incomplete triaging and responsible disclosure.
Instead, we present the critical components in our vulnerability discovery pipeline: (1) A static analysis component that generates vulnerability hypotheses using decompilation output, and (2) a verification tool that uses dynamic symbolic execution (DSE) to eliminate false positive vulnerability hypotheses.
We focus on finding two types of vulnerabilities in Windows drivers: buffer overflows, out-of-bound memory accesses, and arbitrary memory accesses.

\subsubsection{Hypothesis Generation using Decompilation}

We exposed AI agents to angr decompiler~\cite{angrsok2016} via its MCP support, and directed them to accomplish the following tasks after loading the target Windows driver into angr:

\begin{enumerate}
  \item Identify where the driver may accept user input (e.g., via IOCTL handlers, networking, or files).
    We will then suggest the locations that agents should focus on for vulnerability hypothesis generation as well as the types of vulnerabilities that we are interested in (e.g., buffer overflows, use-after-frees, local privilege escalation, missing input validation, etc.).

  \item Critically, we ask agents to explore the binary using MCP, understand its implemented features from a high level, and hypothesize potential vulnerabilities that may be present in the code without examining the actual implementation.
    The report that this step generates helps human analysts develop mental models of the binary and identify potential attack surfaces.
    More importantly, when feeding the output of this step to agents for vulnerability hypothesis generation, they seem to generate vulnerability hypotheses that are more abstract.
    Otherwise, agents tend to overly focus on specific code patterns that are prevalent in decompilation output (e.g., pointer accesses of primitive types or array accesses using index variables) and sometimes totally ignore the high-level functionality of the analysis target.

  \item Generate a transitive closure of functions that are reachable from the focus location, give the list to agents, and ask them to generate vulnerability hypotheses using the decompilation of these functions.
\end{enumerate}

\medskip
\noindent
\textbf{Decompilation quality matters.}
The quality of decompilation output is critical for the quality of the vulnerability hypotheses that agents generate.
Empirically, we find the following types of information are particularly important for agents in vulnerability discovery:

\begin{itemize}
  \item
    The type information of stack variables, global variables, and function arguments.
    Specifically, the differentiation between primitive pointer types and struct pointer types is critical; correctly identifying a struct pointer (and properly typing all its subsequent uses) substantially reduces the number of false vulnerability hypotheses that agents generate regarding the use of this pointer.
    Another important improvement is the recovery of struct types for data structures that are widely used in the analysis target, e.g., a global struct or a pointer to a struct that is passed around functions;
    This improvement also reduces the number of false vulnerability hypotheses that agents generate.
    To support this task, we implemented full-binary type inference in angr decompiler and ingested angr decompiler with Windows kernel-specific type and function definitions.

  \item
    The names of functions, local variables, and global variables.
    Agents generate better vulnerability hypotheses when the decompilation output contains meaningful names.
    Therefore, we explicitly conduct LLM-assisted variable renaming and then function renaming after decompilation to enhance decompilation output.

\end{itemize}

\subsubsection{Hypothesis Verification using DSE}

Empirically, we find that most vulnerability hypotheses that agents generate are false positives. 
As of now, a human is still needed to triage the vulnerability hypotheses that agents generate.
Therefore, quickly eliminating false positives out of these hypotheses is critical to not overwhelm the human analysts.

Without resorting to the hard problem of rehosting and running arbitrary Windows drivers in a test environment, we relied on DSE (in angr) for hypothesis verification.
Specifically, given a vulnerability hypothesis, we task agents to:

\begin{itemize}
  \item
    Identify dynamically generated memory values that may be relevant to exercising the vulnerable code path (e.g., data in runtime-initialized memory regions and pointers in runtime-initialized function pointer arrays). 
    This is to help DSE so that the execution is as accurate as possible if it relies on any dynamically generated values in memory.

  \item
    Generate an angr script that exercises the vulnerable code path, limited to up to three layers of function calls for scalability concerns.
    Then execute the generated angr script to check if the vulnerable code path is reachable and the hypothesized vulnerability condition holds under some input.
    This step eliminates a lot of vulnerability hypotheses that simply cannot hold (e.g., the hypothesis ignores bound checks that happened in a callee function).

  \item
    Generate an input case that may trigger the hypothesized vulnerability then use DSE to check if the generated input case can trigger the hypothesized vulnerability.
    This step rarely works, but it directly leads to a high-likelihood vulnerability finding when it works.
\end{itemize}

\subsection{Firmware}\label{sec:domain-firmware}

We used AI agents to discover vulnerabilities in small yet real-world firmware for embedded systems.
Because these findings are under coordinated disclosure, we do not reveal the affected products or the details of the vulnerabilities.
Instead, we describe the discovery procedure, the role LLM agents played within it, and the limitations we observed while leveraging agents for vulnerability discovery on real-world firmware.

\subsubsection{Hypothesis Generation}
Our vulnerability research on embedded system firmware follows the steps below.
Before the analysis, we set up binary analysis tools with their MCP for the agent to use.  
During the analysis, we prompted the agent to read the firmware image and identify the open-source projects it most likely incorporates, such as network stacks, parsers, and protocol libraries that are widely reused across embedded products.
We then asked the agent whether those upstream projects carry known vulnerabilities and whether the same flaws are still present in the firmware, which often ships a frozen or lightly modified fork that lags behind upstream security fixes.
Beyond known issues, the agent analyzes the corresponding source code to hypothesize previously unknown vulnerabilities, then maps those hypotheses back to the firmware to check that the vulnerable code is actually reachable in the deployed image.
For each candidate, we required the agent to propose concrete validation steps.

\subsubsection{Hypothesis Validation}

For validation, we did not provide concrete instructions; instead, we asked the agent to come up with a plan to validate the identified vulnerabilities, and we allowed the agent to request human assistance in its validation.
Interestingly, the agent asked the human not only to run commands or install necessary tools, but also to purchase devices for vulnerability validation.
For example, the agent asked human researchers to purchase a Bluetooth sniffer in order to validate Bluetooth communication issues in the firmware.
It even provided specific brands and models based on the consideration of available resources and the ecosystem associated with the sniffer device. 
Furthermore, the agent will require the human to physically operate on the Bluetooth devices so that the agent can flash the device with a sniffer image.
Such validations make humans necessary in the loop of vulnerability research.
A vulnerability researcher supplies the needed harness to exercise those steps, either through emulation or on a physical device, and the agent then iterates on validation, refining or discarding hypotheses based on the observed behavior and pausing to request human assistance when it gets stuck.

\subsubsection{AI Agent Effectiveness Impression}
We observed that agents, like a static analysis tool, could not pinpoint the vulnerabilities correctly without human operators' requirements on validation.
One example is the BLE vulnerability as shown in Listing~\ref{lst:micropython}.
The agent identified the code and reported it as a BLE encryption vulnerability, as the firmware image contains constant and blank ER/IR keys for BLE communication.
But during iterative validation prompted by human operator, it recognized from the device's behavior that its discovery was wrong.
Through the observation of the captured Bluetooth traffic, it realized that the traffic was not encrypted and that the configuration of the firmware does not activate this piece of code.
The agent managed to revise it, rather than continuing to assert a finding that the evidence did not support.

\begin{listing}[tb]
\input{sections/micropython.pygtex}
\caption{A code snippet that the agent initially reported as a vulnerability, but was later found to be a false positive.}
\label{lst:micropython}
\end{listing}

On the other hand, the naive workflow, in which we directly ask the agent to point out vulnerabilities in the firmware, is markedly less effective: without the open-source anchor and a validation loop, the agent produces ungrounded guesses that are difficult to confirm and frequently spurious.
Also, without binary reverse engineering tools' support such as IDA Pro MCP, we observed that the agent is less effective in identifying vulnerabilities due to its context limit.

During firmware analysis, we encountered an insufficient context window.
Besides using memory and prompting the agent to documentation, we asked the agent to create specific skills related to different firmware components, such as a Bluetooth security researcher and a GPS security researcher etc, which mirrors real-world human vulnerability research scenarios where different researchers often specialize in different subdomains.
This specialization helps mitigate context window limitations because hypothesis generation and validation typically require multiple rounds of iterative reasoning.
By assigning each subagent to a specific domain, we ensure that its context remains focused on information relevant to the current hypothesis, rather than being diluted by unrelated details from other firmware components.
Furthermore, specialized subagents complement progress documentation. In practice, we observed that agents are not always effective at consistently updating documentation or maintaining coherence across documents over time, which perhaps also results from an insufficient context window.
By encapsulating domain-specific knowledge and reasoning within dedicated subagents, we reduce reliance on continually maintaining comprehensive documentation while preserving focused context for long-running investigations.

\subsection{Web Applications}\label{sec:domain-web}

We identified several critical vulnerabilities in web applications.
We refrain from naming these applications or providing vulnerability details as they are still being fixed by the vendors.
We discuss our experience in verifying vulnerabilities that the agents reported by following the workflow described in Section~\ref{sec:workflow}.

Agents reported more than 200 vulnerabilities across approximately 25 web applications, each accompanied with a PoC.
Agents verified these vulnerabilities by running PoCs against the local deployment of the website (which was also set up by the agents).
Given the volume of reports, we ran another agentic triage workflow to narrow the scope to remote-attacker-controlled vulnerabilities.
After agentic triaging, we were left with half the reports to sift through manually.

Each report generally contains a description of the bug, a list of affected files and lines, a PoC, and a potential patch or recommendations to mitigate the vulnerability.
For manual verification, we started by reading the reports and deploying the website using Dockerfiles or artifacts generated by agents.
Then we did reconnaissance to understand the intended use and functionality of the application.

We used a combination of different verification approaches.
We ran the agent-generated PoC against the deployed web application to confirm whether it worked as expected and to observe its impact.
We reviewed the vulnerable codebase, the PoC or any available logs, and optionally manually reproduced the vulnerability by interacting with the web application.
For some vulnerabilities, we patched vulnerable code and re-ran the PoC to see if the vulnerability was mitigated.

The analysis gave us insights about the capabilities and limitations of agents.
It is interesting that there were many false positives despite our use of dynamic analysis for verification of PoCs.
This was mainly for two reasons.
(1) We observed that agents flagged vulnerabilities that were sometimes only exploitable in a test environment.
Agents would miss some key details such as environment variables or conditions that distinguish production from test environments.
For example, a GitHub Actions workflow would expose the AWS credentials of several critical applications when triggered.
But this workflow can only be triggered by a developer who already has access to the organization's GitHub credentials.
Such cases can be mitigated by refining vulnerability scoping and setting up environments that mirror production.

(2) Agents may generate vulnerability reports (false positive alarms) that do not hold under the context of a specific target.
For example, a third-party application had OAuth access to User A and User B.
The agent hypothesized that because of this OAuth access User A could now have access to User B's data.
But the agent missed a tiny detail that per business logic both User A and User B have access to the same data.

Agents are suited for logic vulnerabilities, as they stem from fine-grained violations that are very specific to the application, and agents create custom hypotheses and verify them, which in many cases is a manual effort.
We also noticed that, given a list of vulnerabilities, agents can chain some of the vulnerabilities together to form an end-to-end powerful attack.

\subsection{Agentically-Reimplemented Code}\label{sec:reimplementation}

Large-scale \emph{reimplementation} of code, whether for memory safety (e.g., the rewriting of the \texttt{coreutils} C package into the Rust \texttt{uutils}~\cite{uutils}) or other reasons (such as the reimplementation of the entire Android mobile operating system into Huawei's OpenHarmony~\cite{independentByHuawei}), has increased in popularity in recent years.
Unfortunately, it has been shown through bitter experience that these reimplementations are often susceptible to the same types of bugs that plagued (and were fixed in) the originals~\cite{chen2026sok,uutilsCVEs}.
These targets are interesting for agentic analysis: because we know examples of specific bugs that were found in the original versions of software, we can better target our search to find re-emerged instances of such bugs in the reimplementations.

To make matters more pressing, agentic capabilities have recently increased to the point that large, complex codebases can be fully agentically reimplemented~\cite{safelibs}.
Of course, analogous to human-written code, AI-generated code has been shown to be susceptible to vulnerabilities~\cite{fu2025security}.
This capability has implications across the software industry, but for the purposes of this section, it gives us a controlled dataset to evaluate agentic capabilities in reimplementation bug hunting.

Over the last few months, our \texttt{SafeLibs} project has reimplemented 24 ``load-bearing'' C libraries (such as \texttt{libpng}, \texttt{libgcrypt}, and other libraries commonly shipped in software and commonly vulnerable to memory corruption bugs) in Rust~\cite{safelibs}.
Because agentic reimplementations can produce \emph{idiomatic}, memory-safe Rust, these libraries are memory-safer than their originals while maintaining drop-in compatibility.
Upon further analysis, we found that analogous bug susceptibility in agentic reimplementations as in human reimplementations.
\texttt{SafeLibs} specifically attempts to \emph{avoid} bug reimplementations (e.g., by providing information on prior non-memory CVEs to reimplementation agents with instructions to avoid these bugs), but we found that, on average, a single agentic vulnerability detection pass (also seeded with the specifics of prior bugs to re-find) identifies multiple reimplemented vulnerabilities per library.

This area deserves additional exploration as agentic code reimplementation becomes increasingly commonly used industry-wide.

\section{What Agents Enable}\label{sec:agents}

A persistent bottleneck in automated vulnerability discovery is writing rules, harnesses, specifications, or building an oracle to detect the vulnerabilities automatically.
Even vulnerability classes with well-defined detection mechanisms, such as buffer overflows and null dereferences, become harder to detect reliably in complex systems with multi-component interactions and layers of abstraction.
Some vulnerabilities cannot be easily expressed as rules or automated checks and instead require manual review or domain-specific knowledge.
For instance, a missing authorization check is the absence of code tied to application-specific logic, or a business logic bug may violate intended behavior without producing a sanitizer-detectable crash.
Designing these checks requires time and expertise and therefore does not scale.

Agents enable scalability and autonomy.
As described in Section~\ref{sec:workflow}, agents do reconnaissance, form target-specific hypotheses, and interpret and refine these hypotheses, which makes them different from prior classification-based machine-learning approaches or from using a single interactive LLM.
We highlight a few problems where we have found agents to be effective.

\noindent
\textbf{Input generation.} Fuzzers and symbolic execution are used to generate inputs that reach a vulnerable function.
Fuzzing relies on predefined corpora or handwritten grammar for grammar-based fuzzing, and even after several mutations the input may or may not reach the target functions.
Symbolic execution is limited by the constraints a solver can solve.
Agents do not suffer from these problems: they can infer the code and generate inputs that are structurally sound and can reach the target function by refining the input using execution feedback.

\noindent
\textbf{Auto harnessing.} Writing a harness is a manual and challenging process.
It requires some knowledge of how different components interact, the correct usage of the API, and so on for the project.
As a result, projects remain under-harnessed.
This limits the ability of fuzzers to explore more code, since harnesses are the entry points for fuzzers.
Agents can help alleviate this problem by generating harnesses for the entire codebase.

\noindent
\textbf{Vulnerability hypothesis creation.} Hypothesizing where a vulnerability might exist requires human expertise.
Agents can reason about code and have knowledge of common vulnerability patterns, which can be further augmented with domain-specific knowledge.
Therefore, agents can create hypotheses about potential vulnerabilities in a target codebase.

\noindent
\textbf{Logic bugs and custom sanitizers.} Memory corruption bugs can often be automatically detected by sanitizers such as ASan, MSan, or KASAN, whereas with logic bugs the violation is not a sanitizer-reported crash.
The violation is a deviation from intended behavior, so logic bugs require manually added annotations or custom sanitizers for automatic detection.
As we discussed in Section~\ref{sec:domain-firmware}, agents are promising in this case.
Agents can generate application-specific oracles or invariants that make logic violations observable.

\noindent
\textbf{PoC generation.} PoCs are used to verify vulnerability.
It is a tedious and iterative process to write a successfully working PoC.
Some PoCs are harder than others, as they may require setting up the target environment or interaction in a specific way to trigger vulnerability.
For instance, as discussed in Section~\ref{sec:domain-xnu}, the agent wrote custom implementation of the network protocol because the off-the-shelf implementation was not directly usable in our case.

\noindent
\textbf{Continuous decision-making.} Vulnerability discovery is a multi-step and iterative process; the next step often depends on the outcome of the previous one.
Human analysts evaluate intermediate results, decide on the next action and adapt when an approach fails.
All of this is continuous decision-making.
Agentic vulnerability discovery enables adaptive decision-making, which wasn't possible before.

For instance, the process of finding vulnerabilities in embedded firmware generally requires a human analyst to identify the correct architecture of the firmware board, determine the peripherals needed to emulate the configuration correctly, or identify open-source libraries used in this embedded firmware.
All of these steps are often manual and do not scale.
In contrast, an agent can autonomously find the correct emulation configuration and set up an environment for the analyst.
Similarly, fuzzers produce hundreds of crashes, which analysts often have to triage and deduplicate before reporting.
Agents can make this process automatic and scalable.

\section{What Now}\label{sec:whatnow}

The vulnerability research landscape is rapidly evolving as agents are finding vulnerabilities in real-world software.
In this section we outline some of the research challenges that we think are important.
\newline

\noindent\textbf{Verification.} Agents report a large number of vulnerabilities and it is infeasible to verify them manually.
Therefore, it is important to develop automated verification techniques to distinguish real vulnerabilities from spurious findings.
There are many techniques such as DSE, fuzzing, or even agents themselves, that can be used for validation of hypotheses.
Agents have been effective in writing PoCs and setting up automated validation pipelines, as discussed in Sections~\ref{sec:domain-web},~\ref{sec:domain-xnu}, and ~\ref{sec:domain-firmware}.

In our experience, dynamic verification has been effective at filtering out false positives.
For traditional software and operating systems, candidate findings can be verified by running or instrumenting the software.
In contrast, firmware can either be emulated, which requires substantial rehosting effort, or validated on physical hardware, which is slow and limited in availability.
We see emulation itself as a promising place for agents to contribute next.

\noindent
\textbf{Measurement.} As models' capabilities advance, we need to revisit the open challenges in vulnerability research.
Tasks that once required deep semantic understanding, such as grammar-based fuzzing, structured input generation, and automatic harnessing, are now far more approachable.
We should measure where agents outperform existing approaches, how they complement them, and where they fail.

\noindent\textbf{Observability.}
Agentic systems are obscure.
Agents invoke tool calls, make plans, have context, and save interactions in memory.
All of these influence the decisions agents make without providing the user with much visibility into them.
For instance, an agent may come up with ten hypotheses and discard some of them without validating them, or it may be biased toward one.
This could be because of an outdated document in the repository, old saved memory, or losing context.
The discarded hypothesis could be a very critical vulnerability.
Therefore, we should develop techniques that give visibility into decision-making or intermediate results that the user can inspect.

\noindent\textbf{Supply chain.}
Agents are used to vibe code software or do vulnerability analysis, often using MCP servers, skills, or plugins present in the agentic ecosystem.
In many cases, the user is unaware of the libraries an agent imports and installs.
This creates supply chain risk because agents can introduce compromised and untrusted components into the workflow and software development.
Future research should develop techniques to automatically inspect externally introduced components.

\noindent\textbf{Scalability and Cost.} With AI, new code is generated at an unprecedented rate, so our vulnerability discovery techniques must scale.
Using agents to find vulnerabilities is one way to scale.
But agentic workflows use large numbers of tokens, and tokens are expensive (when using a frontier model).
Researchers should develop techniques to make agentic workflows token-efficient and build effective open-source models for vulnerability discovery.

\noindent\textbf{Cybersecurity guardrails.} Current cybersecurity guardrails in frontier models hamper cybersecurity defense research and capabilities, even for cyber-verified researchers.
Defensive research requires understanding of offensive techniques to design effective vulnerability mitigations, but guardrails can interfere when researchers study public offensive-related content (novel vulnerabilities and attack techniques).
In some cases, even managing a website containing reported vulnerabilities can trigger these safeguards.
Safeguards can also affect the reliability of findings from autonomous agentic workflows by unexpectedly switching the model or stopping the task.
Therefore, the boundary should be defined more clearly to prevent misuse without obstructing legitimate defensive security research.

\noindent\textbf{Dual-use.} Frontier and open-source models are accessible to both security researchers and adversaries.
Agentic workflows can be used to automate reconnaissance, vulnerability discovery, and exploit development, all at greater scale and speed.
This raises an important research question: how can the security community stay ahead when powerful agentic capabilities are widely available to both attackers and researchers?

\begin{acks}
This project has received funding from the following sources:
National Science Foundation (NSF) Awards No. 2442984, 1663651, 2247954;
Department of Navy Award No. N00014-23-1-2563 issued by the Office of Naval Research;
Air Force Office of Scientific Research Award No. FA9550-24-1-0227;
Defense Advanced Research Projects Agency (DARPA) and Naval Information Warfare Center Pacific (NIWC Pacific) Contract No. N6600122C4026 
and we acknowledge the generous support of the U.S. Department of Defense.
Any opinions, findings, and conclusions or recommendations expressed in this material are those of the author(s) and do not necessarily reflect the views of NSF, DARPA, NIWC Pacific or DoD.




\end{acks}

\bibliographystyle{ACM-Reference-Format}
\bibliography{references}

\end{document}